\newcommand{\MuFourHe}{\ensuremath{\mu{}^{4}\mathrm{He}^{+}}}
\newcommand{\MuThreeHe}{\ensuremath{\mu{}^{3}\mathrm{He}^{+}}}

\newcommand{\Ts}{\ensuremath{\mathbf{Tr_{slow}}}}
\newcommand{\Tf}{\ensuremath{\mathbf{Tr_{fast}}}}
\newcommand{\Te}{\ensuremath{\mathbf{Tr_{elec}}}}

\newcommand{\La}{\ensuremath{\mathbf{L_{\alpha{}}}}}
\newcommand{\Ka}{\ensuremath{\mathbf{K_{\alpha{}}}}}

\newcommand{\Lb}{\ensuremath{\mathbf{L_{\beta{}}}}}
\newcommand{\Kb}{\ensuremath{\mathbf{K_{\beta{}}}}}

\newcommand{\Lg}{\ensuremath{\mathbf{L_{\gamma{}}}}}
\newcommand{\Kg}{\ensuremath{\mathbf{K_{\gamma{}}}}}

\newcommand{\I}{\ensuremath{\mathbf{I}}}
\newcommand{\II}{\ensuremath{\mathbf{II}}}
\newcommand{\III}{\ensuremath{\mathbf{III}}}

\newcommand{\SKa}{\ensuremath{\mathbf{slow~8\,keV}}}
\newcommand{\FKa}{\ensuremath{\mathbf{fast~8\,keV}}}

\newcommand{\slow}{\ensuremath{\mathbf{slow}}}
\newcommand{\fast}{\ensuremath{\mathbf{fast}}}

\newcommand{\TwoSTwoP}{{\ensuremath 2S $\rightarrow{}$2P}}

\documentclass[aip,rsi,reprint,graphicx]{revtex4-1}

\usepackage{xcolor}
\usepackage{textcomp}
\usepackage{graphicx}

\newcommand{\change}[1]{{#1}}

\begin{document}

\pdfinfo {
/Title (Improved X-ray detection and particle identification with avalanche photodiodes)
/Author (Marc Diepold)
}

\title{Improved X-ray detection and particle identification with avalanche photodiodes}

\author{Marc Diepold}
\email[Corresponding author: ]{marc.diepold@mpq.mpg.de}
\affiliation{Max Planck Institute of Quantum Optics,  85748 Garching,
  Germany.}

\author{Luis M.~P.\ Fernandes}
\affiliation{LIBPhys, Physics Department, Universidade de Coimbra, \mbox{3004-516 Coimbra, Portugal.}}

\author{Jorge Machado}
\affiliation{Laborat\'{o}rio de Instrumenta\c{c}\~{a}o, Engenharia Biom\'{e}dica e F{\'\i}sica da
Radia\c{c}\~{a}o (LIBPhys-UNL) e Departamento de F{\'\i}sica da Faculdade de
Ci\^encias e Tecnologia da Universidade Nova de Lisboa, Monte da Caparica, 2892-516
Caparica, Portugal.}
\affiliation{Laboratoire Kastler Brossel, UPMC-Sorbonne Universit\'{e}s, CNRS, ENS-PSL Research
University, Coll\`{e}ge de France, 4 place Jussieu, case 74 75005 Paris, France.}

\author{Pedro Amaro}
\affiliation{Laborat\'{o}rio de Instrumenta\c{c}\~{a}o, Engenharia Biom\'{e}dica e F{\'\i}sica da
Radia\c{c}\~{a}o (LIBPhys-UNL) e Departamento de F{\'\i}sica da Faculdade de
Ci\^encias e Tecnologia da Universidade Nova de Lisboa, Monte da Caparica, 2892-516
Caparica, Portugal.}

\author{Marwan Abdou-Ahmed}
\affiliation{Institut f\"ur Strahlwerkzeuge, Universit\"at Stuttgart, 70569
  Stuttgart, Germany.}

\author{Fernando D.\ Amaro} 
\affiliation{LIBPhys, Physics Department, Universidade de Coimbra, \mbox{3004-516 Coimbra, Portugal.}}

\author{Aldo Antognini}
\affiliation{Institute for Particle Physics, ETH Zurich, 8093 Zurich,
  Switzerland.}
\affiliation{Paul Scherrer Institute, 5232 Villigen-PSI, Switzerland.}

\author{Fran\c{c}ois Biraben}
\affiliation{Laboratoire Kastler Brossel, UPMC-Sorbonne Universit\'{e}s, CNRS, ENS-PSL Research
University, Coll\`{e}ge de France, 4 place Jussieu, case 74 75005 Paris, France.}

\author{Tzu-Ling Chen}
\affiliation{Physics Department, National Tsing Hua University, Hsincho 300,
  Taiwan.}

\author{Daniel S. Covita}
\affiliation{i3N, Universidade de  Aveiro, Campus de Santiago, 3810-193 Aveiro, Portugal.}

\author{Andreas J. Dax}
\affiliation{Paul Scherrer Institute, 5232 Villigen-PSI, Switzerland.}

\author{Beatrice Franke}
\affiliation{Max Planck Institute of Quantum Optics,  85748 Garching,
  Germany.}

\author{Sandrine Galtier}
\affiliation{Laboratoire Kastler Brossel, UPMC-Sorbonne Universit\'{e}s, CNRS, ENS-PSL Research
University, Coll\`{e}ge de France, 4 place Jussieu, case 74 75005 Paris, France.}

\author{Andrea L.\ Gouvea}
\affiliation{LIBPhys, Physics Department, Universidade de Coimbra, \mbox{3004-516 Coimbra, Portugal.}}

\author{Johannes G\"otzfried}
\affiliation{Max Planck Institute of Quantum Optics,  85748 Garching,
  Germany.}

\author{Thomas Graf}
\affiliation{Institut f\"ur Strahlwerkzeuge, Universit\"at Stuttgart, 70569
  Stuttgart, Germany.}

\author{Theodor~W.~H\"ansch}
\email[Also at: ]{Ludwig-Maximilians-Universit\"at, 80539 Munich, Germany.}
\affiliation{Max Planck Institute of Quantum Optics,  85748 Garching,
  Germany.}

\author{Malte Hildebrandt}
\affiliation{Paul Scherrer Institute, 5232 Villigen-PSI, Switzerland.}

\author{Paul Indelicato}
\affiliation{Laboratoire Kastler Brossel, UPMC-Sorbonne Universit\'{e}s, CNRS, ENS-PSL Research
University, Coll\`{e}ge de France, 4 place Jussieu, case 74 75005 Paris, France.}

\author{Lucile Julien}
\affiliation{Laboratoire Kastler Brossel, UPMC-Sorbonne Universit\'{e}s, CNRS, ENS-PSL Research
University, Coll\`{e}ge de France, 4 place Jussieu, case 74 75005 Paris, France.}

\author{Klaus Kirch}
\affiliation{Institute for Particle Physics, ETH Zurich, 8093 Zurich,
  Switzerland.}
\affiliation{Paul Scherrer Institute, 5232 Villigen-PSI, Switzerland.}

\author{Andreas Knecht}
\affiliation{Paul Scherrer Institute, 5232 Villigen-PSI, Switzerland.}

\author{Franz Kottmann}
\affiliation{Institute for Particle Physics, ETH Zurich, 8093 Zurich,
  Switzerland.}

\author{Julian J. Krauth}
\affiliation{Max Planck Institute of Quantum Optics,  85748 Garching,
  Germany.}

\author{Yi-Wei Liu}
\affiliation{Physics Department, National Tsing Hua University, Hsincho 300,
  Taiwan.}

\author{Cristina M.~B.\ Monteiro}
\affiliation{LIBPhys, Physics Department, Universidade de Coimbra, \mbox{3004-516 Coimbra, Portugal.}}

\author{Fran\c{c}oise Mulhauser}
\affiliation{Max Planck Institute of Quantum Optics,  85748 Garching, Germany.}

\author{Boris Naar}
\affiliation{Institute for Particle Physics, ETH Zurich, 8093 Zurich,
  Switzerland.}

\author{Tobias Nebel}
\affiliation{Max Planck Institute of Quantum Optics,  85748 Garching, Germany.}

\author{Fran\c{c}ois Nez}
\affiliation{Laboratoire Kastler Brossel, UPMC-Sorbonne Universit\'{e}s, CNRS, ENS-PSL Research
University, Coll\`{e}ge de France, 4 place Jussieu, case 74 75005 Paris, France.}

\author{Jos\'{e} Paulo Santos}
\affiliation{Laborat\'{o}rio de Instrumenta\c{c}\~{a}o, Engenharia Biom\'{e}dica e F{\'\i}sica da
Radia\c{c}\~{a}o (LIBPhys-UNL) e Departamento de F{\'\i}sica da Faculdade de
Ci\^encias e Tecnologia da Universidade Nova de Lisboa, Monte da Caparica, 2892-516
Caparica, Portugal.}

\author{Joaquim M.~F.\ dos Santos}
\affiliation{LIBPhys, Physics Department, Universidade de Coimbra, \mbox{3004-516 Coimbra, Portugal.}}

\author{Karsten Schuhmann}
\affiliation{Institute for Particle Physics, ETH Zurich, 8093 Zurich,
  Switzerland.}
\affiliation{Paul Scherrer Institute, 5232 Villigen-PSI, Switzerland.}

\author{Csilla I. Szabo}
\email[Currently at: ]{Theiss Research, 92037 La Jolla, CA USA}
\affiliation{Laboratoire Kastler Brossel, UPMC-Sorbonne Universit\'{e}s, CNRS, ENS-PSL Research
University, Coll\`{e}ge de France, 4 place Jussieu, case 74 75005 Paris, France.}

\author{David Taqqu}
\affiliation{Institute for Particle Physics, ETH Zurich, 8093 Zurich,
  Switzerland.}

\author{Jo\~{a}o F. C. A. Veloso}
\affiliation{i3N, Universidade de  Aveiro, Campus de Santiago, 3810-193 Aveiro, Portugal.}

\author{Andreas Voss}
\affiliation{Institut f\"ur Strahlwerkzeuge, Universit\"at Stuttgart, 70569
  Stuttgart, Germany.}

\author{Birgit Weichelt}
\affiliation{Institut f\"ur Strahlwerkzeuge, Universit\"at Stuttgart, 70569
  Stuttgart, Germany.}

\author{Randolf Pohl}
\affiliation{Max Planck Institute of Quantum Optics,  85748 Garching, Germany.}

\date{\today}

\begin{abstract}

Avalanche photodiodes are commonly used as detectors for low energy x-rays.
In this work we report on a fitting technique
used to account for different detector responses resulting from photo absorption in
the various APD layers.  
The use of this technique results in an improvement of the energy resolution at 8.2\,keV by up to a factor of 2, 
and corrects the timing information by up to 25\,ns to account for space dependent 
electron drift time. 
In addition, this waveform analysis is used for particle identification,
e.g. to distinguish between x-rays and MeV electrons in our experiment.

\end{abstract}

\pacs{}

\maketitle 

\section{Introduction}
\label{sec:intro}

\begin{figure}[t]
\begin{center}
\includegraphics[angle=0,width=8.8cm]{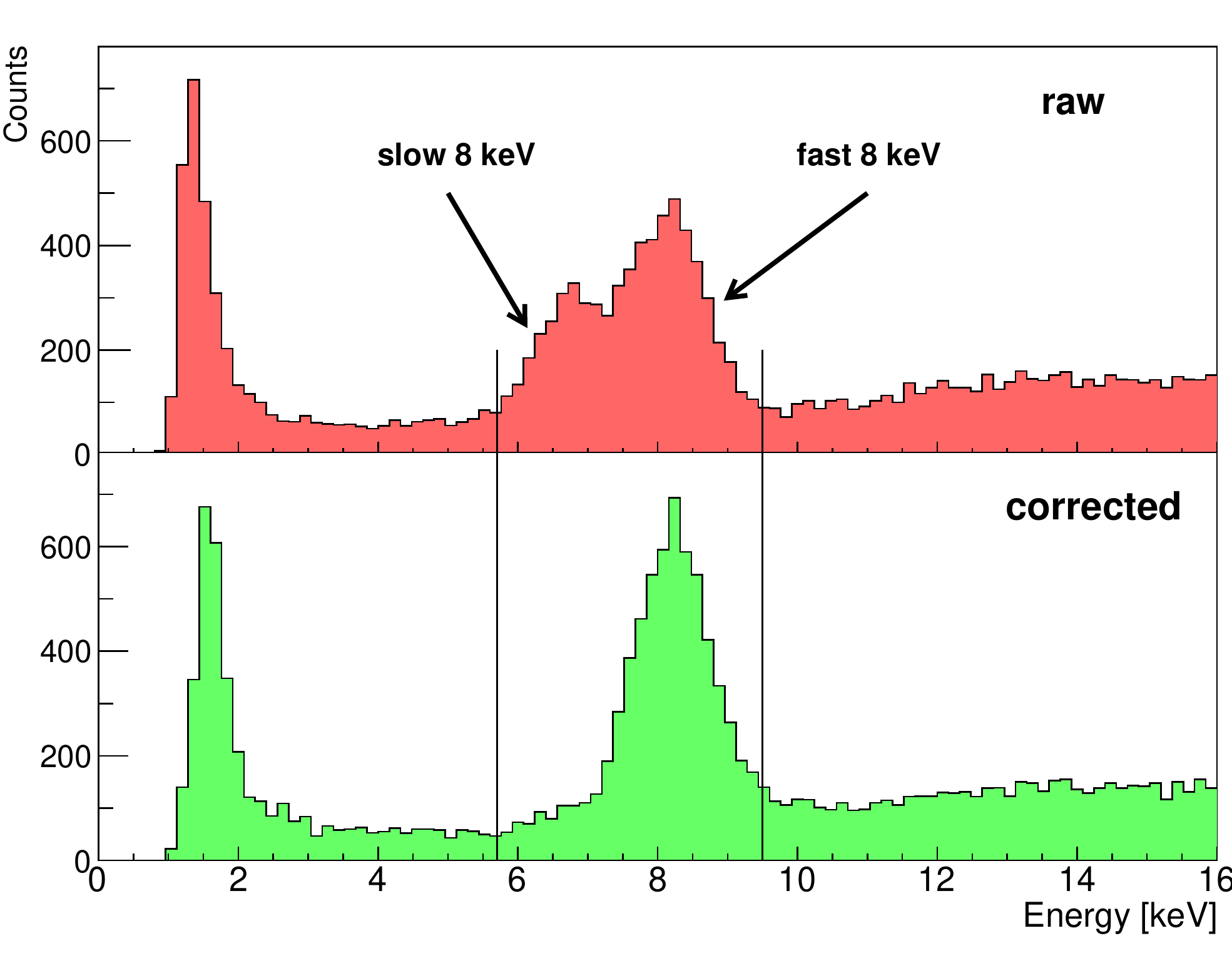}
\caption[X-Ray energy calibration]
{
 X-ray energy spectrum from a single APD before
 (top) and after (bottom) applying our correction. The first spectrum is obtained by
 integrating over the recorded pulse amplitude in a 200\,ns time window after the leading
 edge. A difference in extracted energy for \change{detector responses with slow and fast rise times (labeled \SKa{} and \FKa{} x-rays respectively)} is 
 clearly visible. Improved energy calibration managed to unite both responses and
 improve the energy resolution by up to a factor of 2 (from 32\%{} to 16\%{}\,FWHM at
 8.2\,keV for this APD). 
}
\label{fig:ecal}
\end{center}
\end{figure}

Avalanche photodiodes (APDs) are silicon-based solid state detectors that
convert photons into a charge current. 
They provide a compact, robust, magnetic field insensitive
solution for light and x-ray detection with gains on the order of 100 and fast
response times~\cite{apd_stuff, apd_stuff2, pancho, pancho2}.  
Due to this, APDs are extensively used in a large variety of physics~\cite{bla1,bla2,bla3,bla4}, medical~\cite{bla5}
and aerospace applications~\cite{bla6}.  
We have studied x-rays with energies between 1 and 10\,keV and observed two distinct APD responses to monoenergetic
x-rays absorbed in different depths inside the APD.
By constructing APD specific standard traces, and using a pulse-by-pulse
fitting technique, we improved the APD energy resolution by a factor of 2, and
the time resolution by 30\%.
In addition, we were able to identify background signals stemming from 
electrons that deposit a few keV energy in the APD.
The data presented in this work were gathered in the muonic helium
Lamb shift experiment~\cite{exp_muhe1,exp_muhe2}, using a set of twenty large area
avalanche photo diodes (LAAPDs) from Radiation Monitoring Devices (model
S1315; $13.5 \times 13.5$\,mm$^2$ active surface area each).
The muonic helium ions represent an extended x-ray source that emits predominantly monoenergetic x-rays of
1.52\,keV and 8.22\,keV as well as electrons with up to 50\,MeV of kinetic energy
(see Appendix~\ref{app:exp}).
Previous tests of these APDs found 40\%{} detection efficiency for 8.2\,keV x-rays,
and an average energy resolution of 16\% (FWHM) after calibration~\cite{old_apd}.  
Our x-ray detection setup consists of two linear arrays of 10 LAAPDs each, in which 
each LAAPD is mounted on a separate titanium piece for efficient cooling and 
easy replacement~\cite{pancho,old_apd}.
The detector arrays are
mounted inside a vacuum around 10$^{-5}$\,hPa, and inside a 5\,Tesla magnetic
field, above and below the x-ray source. 
Custom-built low-noise, fast response preamplifiers are fitted to the LAAPDs.
Both LAAPD/pre-amplifier assemblies are cooled using an external
ethanol circulation system and are actively temperature stabilized at around
$-30^{\circ{}}$C.
The achieved short term temperature stability was better than $\pm{}0.1^{\circ{}}$C.
Highly stable temperatures are crucial for the operation of LAAPDs since their
gain depends strongly on their operating temperature~\cite{pancho,old_apd}.
Bias voltages were chosen to provide the best energy resolution per APD and
ranged from 1.61\,kV up to 1.69\,kV, \change{approximately 50 Volts below the 
breakdown voltage.}
The pre-amplifiers with two bipolar input transistors in cascode configuration
(BFR 182 npn, BFT 92 pnp) have been used for the generation of a fast response
from the large capacitance (120\,pF) of the LAAPD.
An overall gain of 150\,mV/$\mu$A at 50\,$\Omega$ has been measured with a
test pulse.
%
%
Outgoing APD signals were further amplified by gain 4 main-amplifiers and 
fed to the CAEN v1720 waveform digitizers (250 MS/s, 12 bit) 
for recording.

\begin{figure}[tb]
\begin{center}
\includegraphics[angle=0,width=8.8cm]{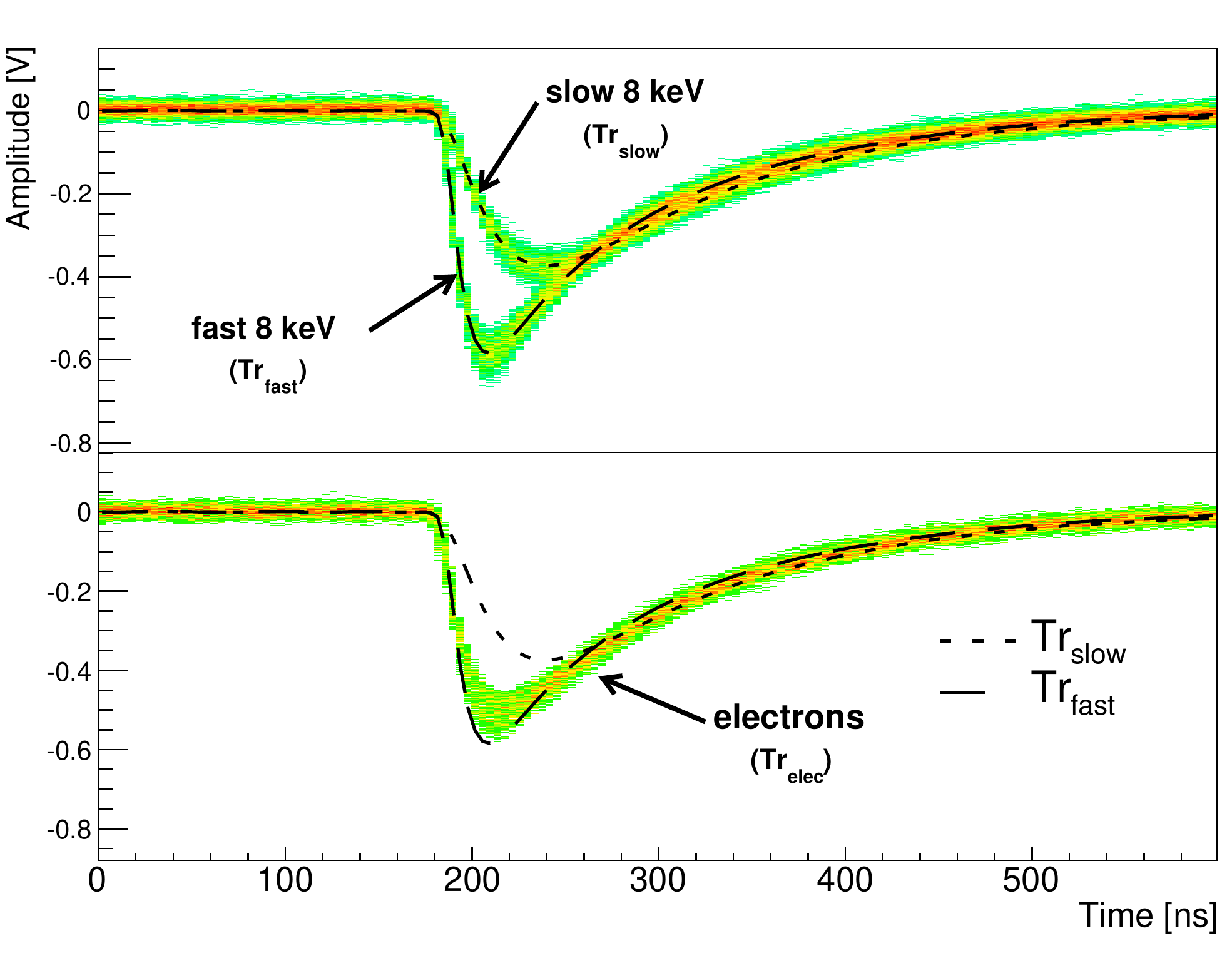}
\caption[APD slope versus integral plot]
{
  Top: Typical APD responses for 8.2\,keV x-rays. Even though
  incoming x-rays are quasi mono energetic, the APDs show two distinct
  responses. 
  The \FKa{} component has a rise time of about 35\,ns while
  the \SKa{} component shows a rise time of about 70\,ns. Separate averaging
  of both individual data sets allows to produce standard traces that accurately describe all
  x-ray traces between 2\,keV and 10\,keV.  Bottom:
  Electron induced signals that correspond to an x-ray energy of 8.2\,keV after
  calibration. The dashed curves show the average of the \SKa{} and
  \FKa{} x-rays.  Even though similar in shape to fast 8\,keV x-ray signals,
  a $\chi{}^2$ fit was able to identify 86\%{} of these electrons
  correctly. 
}
\label{fig:trace}
\end{center}
\end{figure}

Our experiment requires pileup detection in the x-ray detectors to reduce
background effects.
Standard shaping amplifiers that are commonly used
feature integration times too long to 
separate pulses on a 100\,ns scale.
This deteriorated the performance in our previous measurements~\cite{
Pohl:Nature, Antognini:Science, Diepold:PRA} where we used Rutherford Appleton Laboratory (RAL) 
108A pre-amplifiers with $\mu$s-long integration times \change{(see~\cite{old_apd} Fig.~16).}
\change{For our new project~\cite{exp_muhe1,exp_muhe2}} we used fast pre-amplifiers with 30\,ns rise time.
When calculating a simple integral over the recorded pulses, a poor energy resolution 
became visible as seen in Fig.~\ref{fig:ecal}.
The double peak structure \change{that was clearly resolved in 6 out of 20 APDs} is a result of two different APD
responses to the monoenergetic 8.2\,keV x-rays as can be seen in the upper
part of Fig.\,\ref{fig:trace}. 
Similar effects were previously reported for beveled edge APDs and
14.4\,keV x-rays~\cite{apd_baron}.
\change{We first observed the same behaviour in a separate test 
setup without magnetic field.
Hence the features described here can not be attributed to magnetic trapping effects
in the drift region~\cite{apdmag}.
We can only speculate why this effect was not seen for x-rays taken in another experiment
at cryogenic temperatures~\cite{cooper}.
Pre-selection of APDs with good energy resolution at 5.9\,keV can lead to the vanishing of the double 
peak structure.
This was the case in our previous measurement~\cite{old_apd}.
Also the large average angle of incidence in our setup increseases this effect significantly.} 
To compensate, we developed a simple standard response fitting technique that allowed us to distinguish
between different responses on a hit-by-hit basis, improving the energy
resolution by a factor of two (see Fig.\,\ref{fig:ecal}, bottom) and correcting
for a 25\,ns time shift between both signal types as discussed in Section\,\ref{sec:time}. 
In the Secs.~\ref{sec:inter}-\ref{sec:summary}, the different features of the measured x-ray signals are discussed before
the fitting routine and the improved energy calibration are
presented.
Then timing difference between both responses and the influence of
electron signals in the analysis are reviewed before a brief summary and
outlook is given. 

\section{APD x-ray response}
\label{sec:inter}

\begin{figure}[t]
\begin{center}
\includegraphics[angle=0,width=8.8cm]{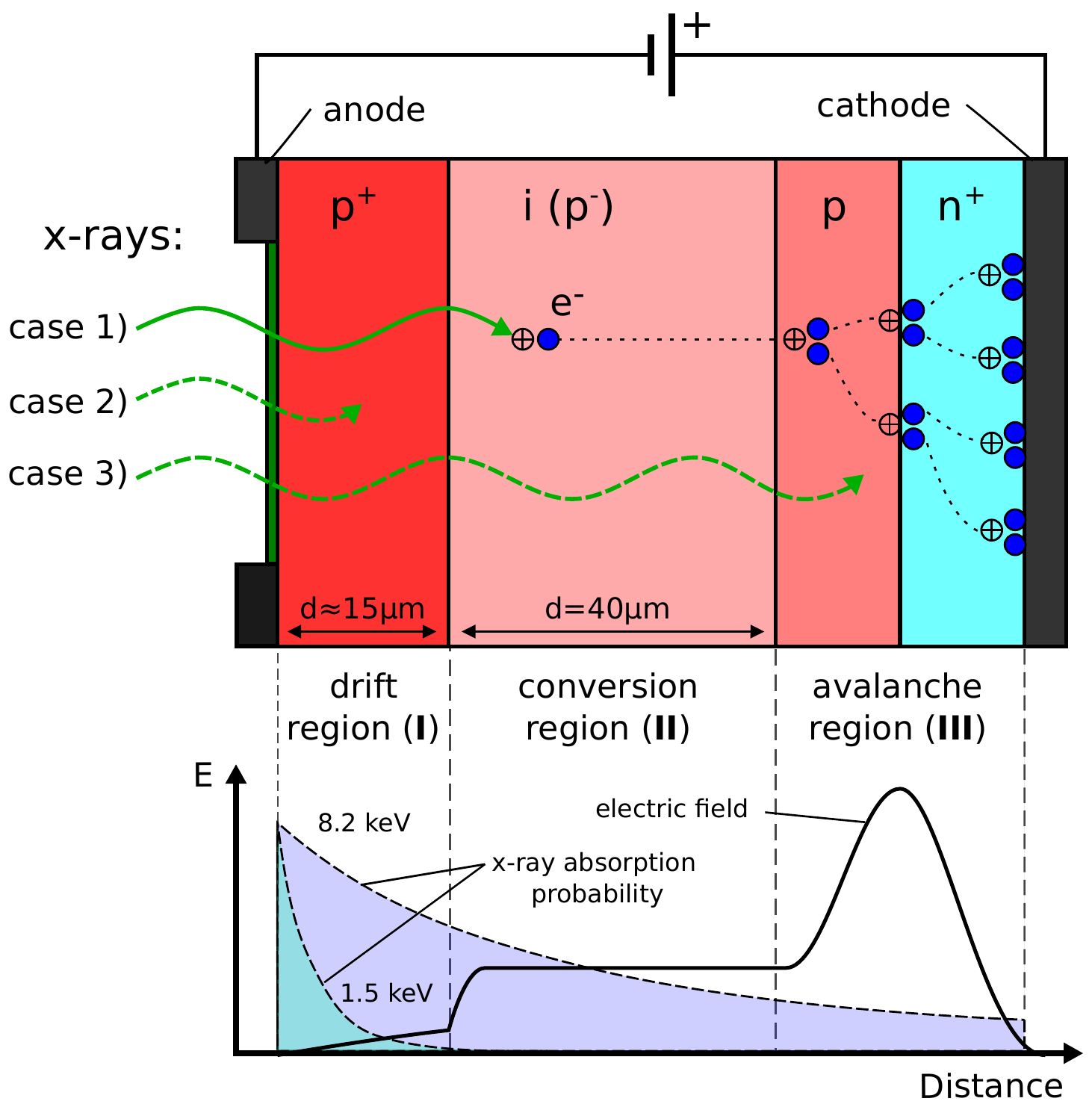}
\caption[APD scheme]
{
  Working principle of avalanche photodiodes based on a 
  p$^{+}$-i-p-n$^{+}$ doping profile. The weakly doped
  intrinsic part (\II{}) serves as conversion region for most incoming x-rays (case
  1). Photoelectrons created are transferred towards the avalanche region. In this
  high field area secondary electrons are generated through impact
  ionization providing charge gain. Low energy x-rays have a high
  probability of being stopped in the initial drift region (\I{})(case 2). 
  These experience additional signal delay and reduced
  gain. Some photons convert in the multiplication
  region (\III{}), also leading to reduced signal amplitudes (case 3).
  More about this effect can be found in~\cite{apd_baron}.
  The bottom figure shows the electric field profile in the several regions of
  the APD together with the x-ray absorption profile for 1.5\,keV and 8.2\,keV x-rays.
}
\label{fig:scheme}
\end{center}
\end{figure}
The working principle of the APDs used in our setup is 
explained in Fig.\,\ref{fig:scheme}.
In the conversion region (\II{}), incoming photons
produce primary photoelectrons.
Differences in the thickness of this layer (\II{}) give rise
to changes in detector energy acceptance.  
A p-n junction is placed on the back side of the active volume creating high local field strengths.
Inside this avalanche region (\III{}) electron impact ionization at the high
field p-n$^+$ junction leads to a multiplication of free charge carriers
providing gain for the initially converted primary photoelectrons\,\cite{apd_stuff2}. 
The calculated absorption length for 8.2\,keV and 1.5\,keV x-rays is 70\,$\mu{}$m
and 9\,$\mu{}$m, respectively\,\cite{abs_length}.
Due to the extended size of our x-ray source, the average incident angle of
52 degrees in our geometry gives rise to an effective 1.6 times longer average path 
inside the APDs. 
This absorption length for 8\,keV x-rays is similar to the APD layer thicknesses and therefore
leads to a number of different effects on the APD output depending on the
region where the photon is absorbed.
The different possibilities are also shown and explained in Fig.\,\ref{fig:scheme}.

The largest part of the recorded 8\,keV x-rays stops in the conversion region (\II{}) and follows the normal APD working 
principle that provides high charge collection efficiency and fast amplification. 
Nevertheless, some x-rays are absorbed either in the drift layer
(\I) or in the avalanche region (\III{}).
The x-rays absorbed in region (\III{}) undergo only partial amplification
resulting in low amplitudes down to zero.
This gain reduction is responsible for the flat energy tails seen in Fig.\,\ref{fig:ecal}.
X-rays absorbed in region (\I{}) generate electrons which are only slowly
transfered to the following region (\II{}) due to the lower field strengths in (\I{}).
Traps in this region may hold electrons for non-negligible times, lengthening
the pulse and causing a reduction in amplitude\,\cite{apd_baron} (see Fig.\,\ref{fig:trace}).  
\change{Similar effects of reduced charge collection efficiency were also 
studied for x-ray energies below the silicon K-edge\,\cite{apd_response}.}
From Fig.\,\ref{fig:trace} we also observe that these x-rays only show a single
amplitude and not a continous distribution up to the one of x-rays absorbed in
region (\II{}).
This indicates that the trapping mechanism occurs at the boundary between
regions (\I{}) and (\II{}).

\section{X-ray energy differences and compensation}
\label{sec:meth}

\begin{figure}[t]
\begin{center}
\includegraphics[angle=0,width=8.8cm]{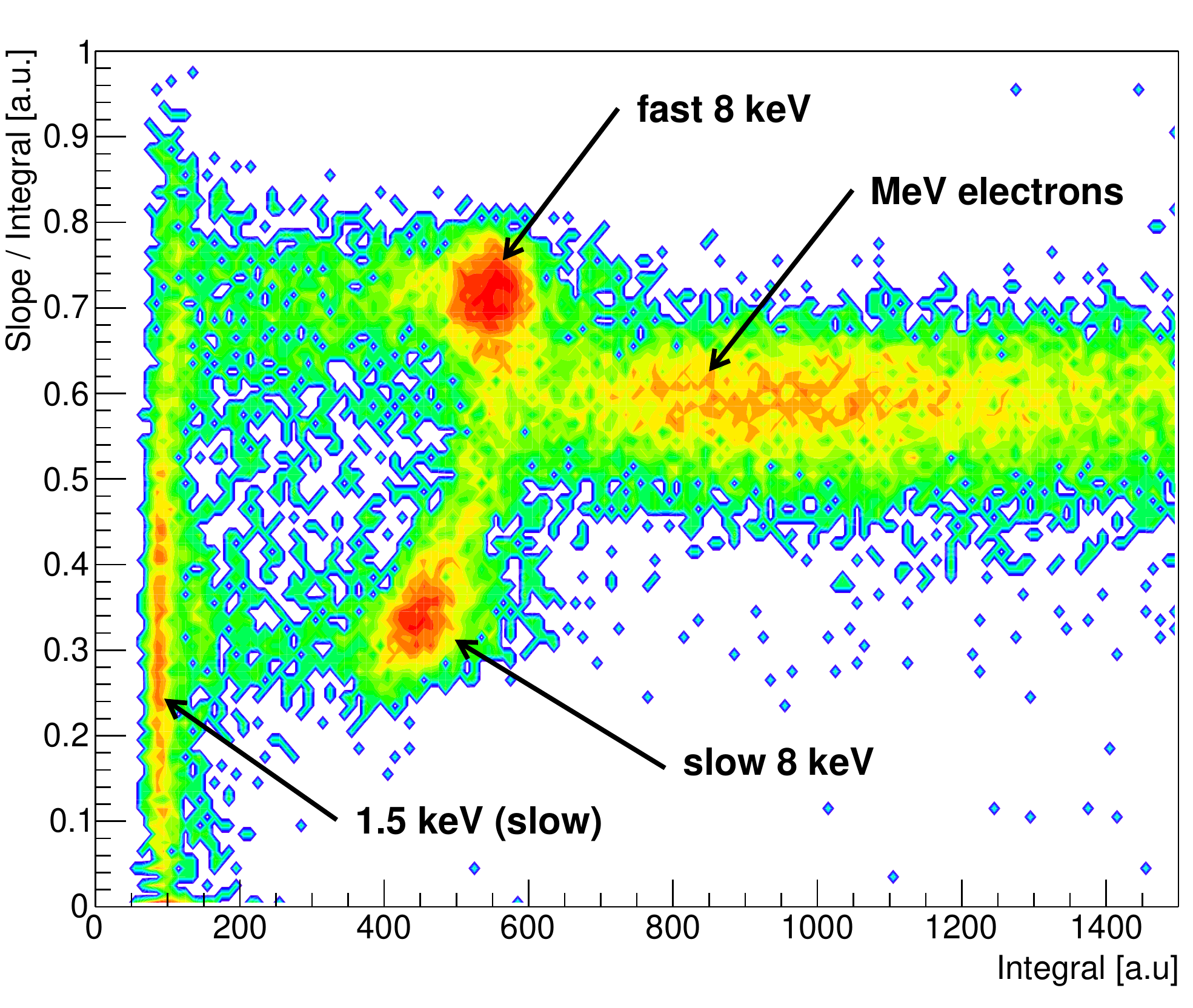}
\caption[APD slope versus integral plot]
{
  Normalized slope of the rising edge plotted
  versus the integral of the pulse. The z-axis~(color scale) is logarithmic.
  Integrals are roughly proportional to the deposited energy of the registered
  x-rays. Four contributions are visible: Low energy 1.5\,keV x-rays show
  integrals below 200. The recorded 8.2\,keV x-rays
  create two different responses in the APD,  
  one with slow rise time (slope $\approx$ 0.3), and one with significantly 
  faster rise time (slope $\approx$ 0.7). The last contribution with an
  integral above 700 arises from MeV electrons depositing keV energy in the 
  APD active region.
}
\label{fig:slopevsint}
\end{center}
\end{figure}

\begin{figure}[tb]
\begin{center}
\includegraphics[angle=0,width=8.8cm]{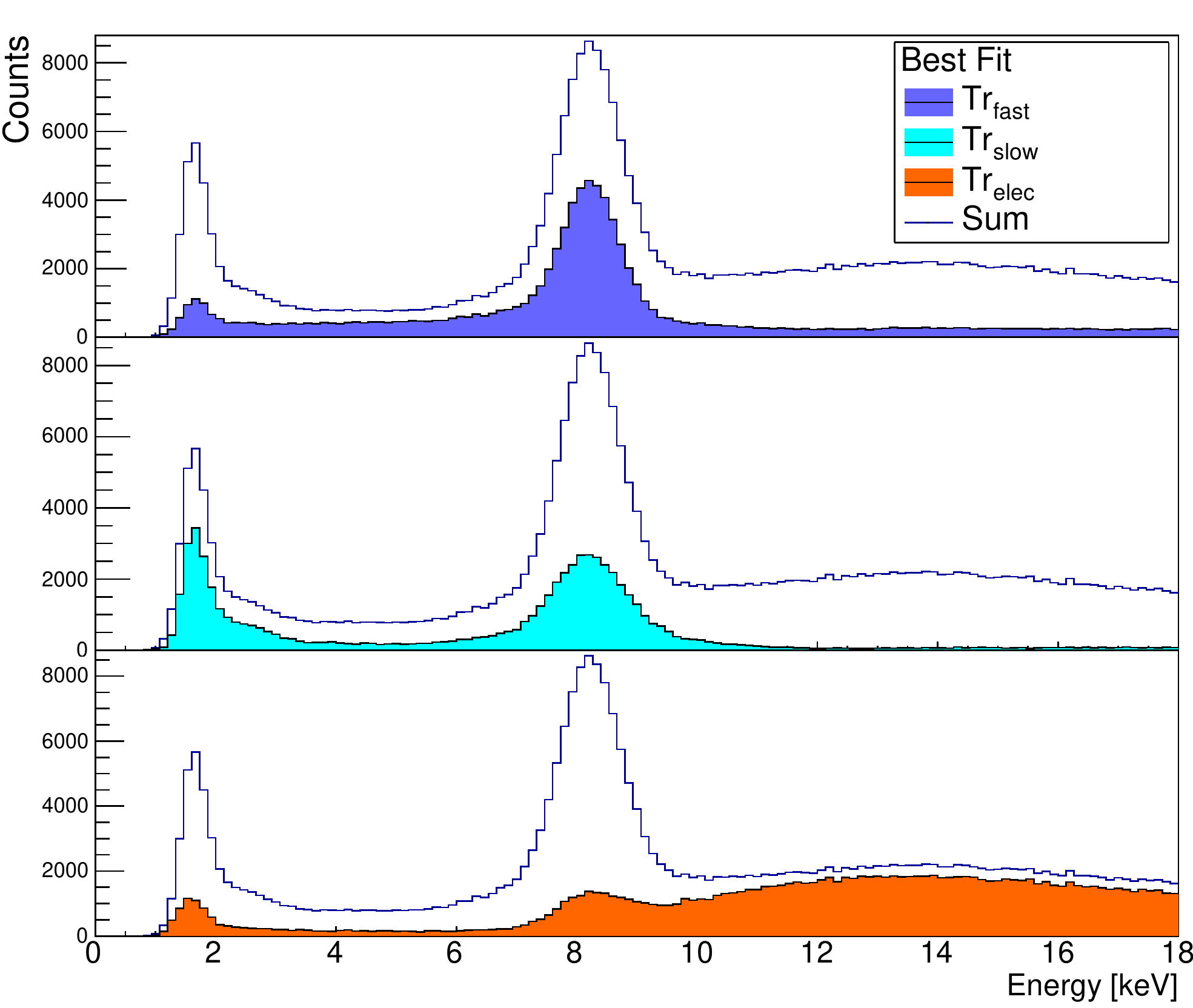}
\caption[APD slope versus integral plot]
{
  Energy spectra of recorded x-rays and electrons in the muonic helium Lamb shift experiment
  categorized by the standard trace which provides the lowest $\chi{}^2$ in a range
  of 200\,ns after the leading edge of the pulse. All spectra show
  two prominent peaks at 1.5\,keV and 8.2\,keV. 
  The fast rising component provided  by~\Tf{} in dark blue enfolds signals
  converted in or behind the conversion region (\II{}). It consists mostly out
  of 8.2\,keV x-rays and the visible low energy tail is created by the loss
  of gain for x-rays converted in the avalanche region (\III{}).
  The light blue distribution
  stands for all traces that were best described by the slow rising pulse shape~\Ts{}
  and consists mostly out of 1.5\,keV x-rays and some 8.2\,keV x-rays mixed in.
  Signals best matching 
  the electron trace~\Te{} are shown in the orange division. These signals are
  formed by a continuous electron background and a contribution of
  wrongly identified x-rays.  
}
\label{fig:energy}
\end{center}
\end{figure}

In order to investigate this effect a set of roughly
$2.5\times{}10^4$ x-ray traces were recorded per APD.
Fitted baseline fluctuations were below 10\,mV for all analyzed
signals, compared to average signal amplitudes of 500\,mV for 8.2\,keV x-rays.
Our analysis routine starts with an
edge finder (square weighting function with a width of 200\,ns) 
to find the beginning of the pulse in the recorded trace.
Then the slope of the leading edge is fitted with a linear function.
Using a $\chi^2$ criterion we improve the accuracy of the slope determination by varying 
start time of the pulse within 20\,ns while keeping the fitting
window fixed.
Finally, we normalize the slope to the pulse integral provided by the edge finder
to obtain the (amplitude-independent) rise time of the pulse.

\begin{figure}[t]
\begin{center}
\includegraphics[angle=0,width=8.8cm]{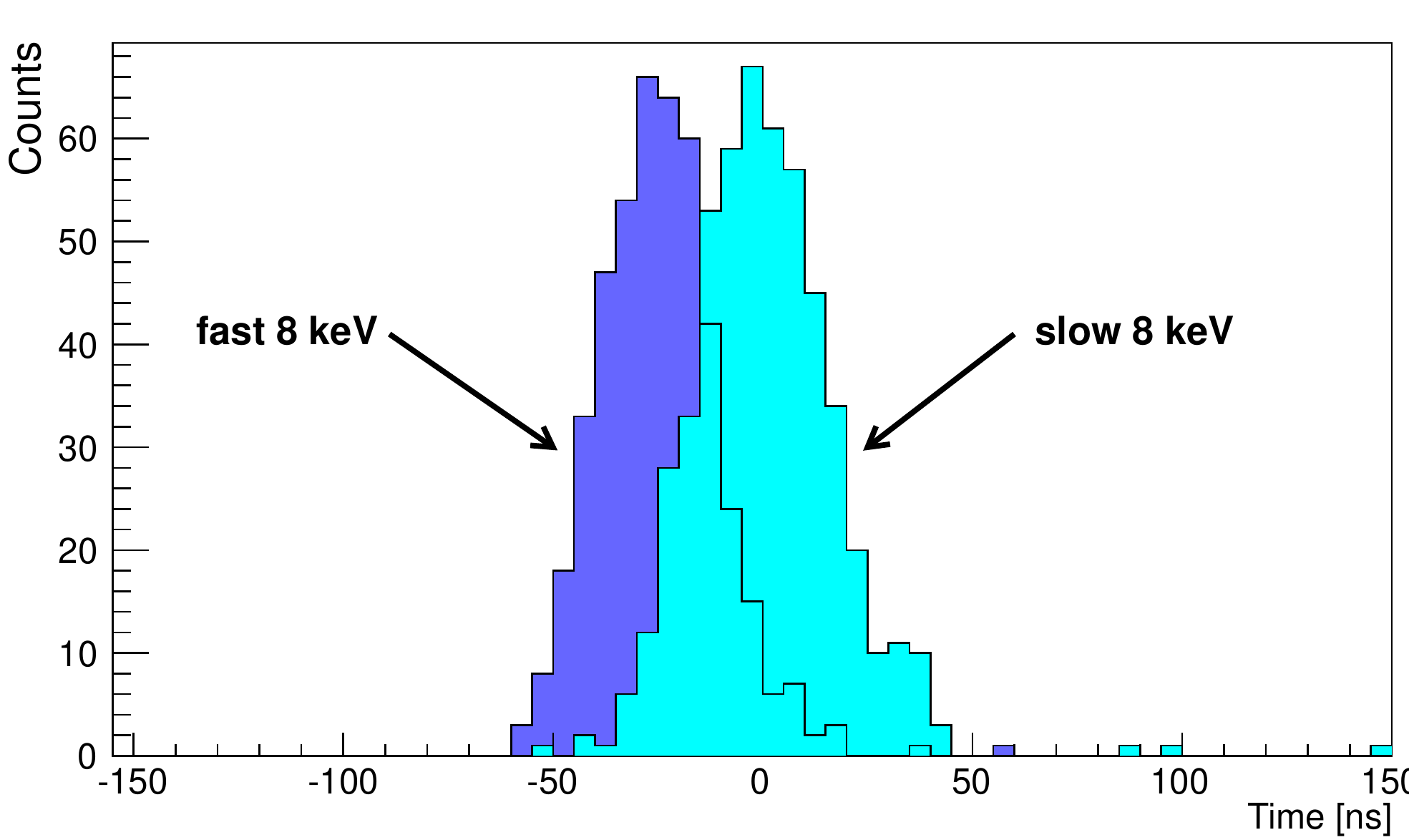}
\caption{Spectra showing the relative timing between \SKa{} and \FKa{}
  x-rays for a single APD. These time spectra have been optained by plotting
  the time of the 8.2\,keV x-ray signal detected in a LAAPD relative to the
  1.5\,keV signal detected in another LAAPD for both classes. Both LAAPDs
  including preamp-delay line etc. have been synchronized using electrons. 
The origin is chosen as the
  center of gravity of the \SKa{} response. A 25\,ns timing difference could be
  measured between both responses. The relatively poor time resolution is
  given by the coincident 1.5\,keV x-rays that are detected just above the
  noise level in our setup leading to a washed out signal. }
\label{fig:time}
\end{center}
\end{figure}

When the rise time is plotted versus the integral of the 
pulse, four different contributions to the spectra can 
be identified as seen in Fig.\,\ref{fig:slopevsint}.
The two most prominent peaks are created by converted 8.2\,keV photons 
\change{with slow and fast detector responses, labeled \SKa{} and \FKa{} respectively}. 
For these peaks we see a clear difference in rise time and integral
while most of the low energy 1.5\,keV x-rays show a slow rise time. 
The rise time distribution for small signals is broadened due to low amplitudes and 
noise.

The last visible component is generated by the already mentioned high energy (up to 50\,MeV) electrons
(created by muon-decay, further explained in the appendix\,\ref{app:exp}).
These electrons deposit energies up to 50\,keV in the APDs and
their signals display a third kind of standard pulse shape, namely a mixture 
of fast and slow x-ray pulse shapes. 
This is shown in the lower panel of Fig.\,\ref{fig:trace}.  
In order to further analyze the two classes of 8.2\,keV x-rays, two sets of APD traces
for \SKa{} and \FKa{} were created by selecting the respective peaks
in Fig.\,\ref{fig:slopevsint} with adequate cuts.
A collection of selected traces for the \SKa{} and
\FKa{} cases is shown in Fig.\,\ref{fig:trace}.
For each of the two x-ray classes, traces were numerically averaged after 
shifting each trace to correct for the variation of the pulse starting time.
This averaging created the standard traces of the subsets (\Ts{} \&{} \Tf{}).
These traces had to be produced once per each APD for a measurement period of several
months and stayed constant throughout multiple heating/cooling cycles of the APD assembly. 
For the final analysis, each APD pulse is fitted with all available standard traces.
Starting at the time provided by the edge finder, 
the standard trace is fitted to the pulse. 
The timing is then varied and the $\chi^2$ is recorded for each fit.
To save computational effort that would arise for a 2-parameter fit (amplitude and time), 
the amplitude of the standard trace is always fixed by matching
its integral to the integral of the signal (after baseline subtraction) in
a 200\,ns wide time window.

Finally, the minimal $\chi^2$ between the various standard traces is used to separate the
pulses into different classes: \SKa{} and \FKa{} (and electrons, see below).
The result from the best-fitting class is used to get amplitude, integral and 
timing values of the recorded signal.

The allocation of the recorded APD signals in the \SKa{} and \FKa{} classes 
according to the fit routine can be seen in the top parts of Fig.\,\ref{fig:energy}.
Calibration of the two x-ray spectra created by the \Ts{} and \Tf{} fits is
done by matching the peaks in both separate integral spectra to the respective
energy of 8.2\,keV.
As expected, the \FKa{} component of x-rays is the largest part of the
recorded signals in our setup as seen in Fig.\,\ref{fig:energy}.
The observed 1:1.7 ratio of \FKa{} to \SKa{} x-rays agrees roughly with the 
expected absorption ratio of 1:1.5 estimated from the thicknesses of layers (\I{})
and (\II{}).

\section{X-ray timing differences}
\label{sec:time}

In addition to the variation in the observed 8.2\,keV x-ray energy 
we were also able to measure a difference in timing between the \FKa{} and the \SKa{}
components.
In order to achieve a common timing reference point for this study,
coincidence events between the 8\,keV x-rays recorded in the APD under investigation and
the 1.5\,keV x-rays registered in neighboring APDs were studied.
These two x-ray types are emitted within a picosecond time window from the
muonic atoms used (as is further explained in the appendix\,\ref{app:exp}). 
Special attention was given to time calibration in order to avoid possible timing
shifts created by the distinct standard traces \Tf{} and \Ts{} 
used for different signals.
Therefore calibration of the different APDs and traces against each other was done using the
supplementary measured electron signals.
The MeV electrons create hits in multiple detectors on their spiraling motion in the surrounding magnetic field enabling us to get a common timing for all APDs.
When comparing the timing of the measured 8.2\,keV x-rays we observed a
25\,ns delay between \SKa{} signals and normal \FKa{} signals.
A time spectrum showing this effect for a single APD is shown in Fig.\,\ref{fig:time}.
Correcting for this effect improves the APD time resolution of our setup by
more than 30\,$\%$ when the two responses seen in Fig.\,\ref{fig:time} are unified.
Better results might be achieved when a more clearly defined common timing
is provided since the timing resolution is limited by the low amplitude 1.5\,keV
signals just above the noise level.

\section{M\MakeLowercase{e}V electron detection with apds}
\label{sec:elec}
Apart from the improved energy resolution that was achieved with the
methods described in Section\,\ref{sec:meth}, we were also able 
to differentiate high energy electron signals in the APDs  
from similar x-ray signals.
These MeV electrons deposit up to 50\,keV in the APD active volume and
were always present in the experiment.
Due to their passage through all the APD layers,
electrons show signals with yet another shape that can be distinguished from the
previously discussed \fast{} and \slow{} x-ray responses.
A third standard trace \Te{} was created by averaging a set of clearly identified electron
signals that correspond to a mean energy of 12 keV.
This was done supplementary to the already known x-ray traces \Ts{} and \Tf{}.
A comparison of an electron induced signal shape at 8.2\,keV and the respective
real x-ray traces is also shown in Fig.\,\ref{fig:trace} (bottom).
As electrons with MeV energies deposit energy in all three APD layers. 
(\I{}, \II{}, \III{}).
The corresponding standard trace \Te{} can be approximately parameterized as a mixture of the 
\Ts{} and \Tf{} standard responses.
Using the same routine as for the previous pulse analysis, the fit was 
able to differentiate between x-ray and electron signals with very high
fidelity, leading to a correct electron identification in 86\%{} of the cases.

\section{Summary}
\label{sec:summary}
We have observed effects from the APD layer structure that lead to two distinct responses
to X-rays in the 6-10\,keV range.
The individual signal types can be identified with high fidelity by examining
the rising edge of the measured pulses.
Correcting for this effect improves the energy resolution
by up to a factor of 2 depending on the APD.
Additionally we were able to correct for timing differences between both responses.
\change{While the different rise time classes were observed in all 20 APDs under
investigation, only 6 of them showed a resolved double-peak structure in
the energy spectrum obtained by a simple integral.}

Using the rise time analysis, it was also
possible to filter MeV energy decay electrons.
An electron-specific standard trace was clearly distinguishable from the two
different kinds of x-ray signals recorded for 8.2\,keV x-rays.
A $\chi^2$ fit of the signal shape was used to exclude them from the x-ray data
with an overall effectiveness of 86\,\%{}, while only 14\,\%{} of the 8\,keV x-rays
were wrongly identified as electrons. 
This lead to significant background reduction in the $\mu{}$He Lamb shift experiment~\cite{exp_muhe1,exp_muhe2}.

\section*{Acknowledgments}
We thank Ulf R\"oser, Matteo N\"ussli, Hanspeter v. Gunten, Werner
Lustermann, Adamo Gendotti, Florian Barchetti, Ben van den Brandt, Paul
Schurter, Michael Horisberger and the MPQ, PSI, ETH Workshops and support
groups for their help.
M.D., B.F., J.J.K., F.M. and R.P.\ acknowledge support from the European Research
Council (ERC) through StG.\ \#279765.
F.D.A., L.M.P.F, A.L.G., C.M.B.M. and J.M.F.S. acknowledge support from FEDER
and FCT in the frame of project PTDC/FIS-NUC/0843/2012. 
C.M.B.M. acknowledges the support of FCT, under Contract No. SFRH/BPD/76842/2011.
F.D.A. acknowledges the support of FCT, under Contract No. SFRH/BPD/74775/2010.
A.A, K.K and K.S acknowledge support from SNF 200021L-138175.
T.G., A.V., B.W and M.A.A. acknowledge support of DFG\_GR\_3172/9-1.
This research was supported in part by Funda\c{c}\~{a}o para a Ci\^{e}ncia e a
Tecnologia (FCT), Portugal, through the projects No. PEstOE/FIS/UI0303/2011
and PTDC/FIS/117606/2010, financed by the European Community Fund FEDER
through the COMPETE. P.~A. and J.~M. acknowledges the support of the FCT,
under Contracts No. SFRH/BPD/92329/2013 and SFRH/BD/52332/2013.

\appendix*
\section{The $\mu{}$He Lamb shift experiment}
\label{app:exp}

The data presented in this work were acquired using muonic helium ions as x-ray
source during the recent $\mu{}$He Lamb shift experiment~\cite{exp_muhe1,exp_muhe2}.
The experiment is performed at the high intensity proton accelerator facility at
Paul Scherrer Institute in Switzerland.
Its' purpose was to measure the different \TwoSTwoP{} transitions in the
\MuFourHe{} and \MuThreeHe{} exotic ions via laser spectroscopy.
The required information about its environment and working principle will 
be briefly sketched in this section.
The accelerator physics environment leads to stringent demands on stability and robustness
of the APDs and the analysis routine employed that exceed common specifications.
For example, the APD arrays used are placed inside a 5\,T
solenoidal magnet where they are mounted next to a low pressure helium gas target.
Muonic ions are created in this 20\,cm long gas volume operated at $2-4$\,hPa by
low energy muons that are provided by the accelerator beam line.
The dataset described in this work was obtained during the \MuFourHe{}
measurement campaign in 2013 that offers multiple transitions in the low keV
x-ray region.
These consist of the \La{}, \Lb{} and \Lg{} transitions at 1.52~keV, 2.05~keV
and 2.30\,keV respectively as well as the \Ka{}, \Kb{} and \Kg{} transitions
at 8.22\,keV, 9.74\,keV and 10.28\,keV, emitted by the muonic
helium ions during the so-called atomic cascade
within a time frame of few ns total~\cite{cascade}.

The muons decay after an average lifetime of 2.2\,$\mu{}$s into muon
neutrino, electron antineutrino and \"{}high energy\"{} electrons in the MeV range.  
These electrons deposit energy when transversing the APD,
creating electron hole pairs in all regions of the APD quasi
simultaneously. 
The induced signals correspond to virtual x-ray energies of up to 50\,keV.
This would raise background effects for the experiment that
uses the recorded 8.2\,keV \MuFourHe{} \Ka{} x-rays as signal for
laser spectroscopy.
Therefore a supplemental set of 4 plastic scintillators surrounds the gas 
target and APD arrays for additional means of electron detection and exclusion
of background.  
Since the overall detection efficiency for electrons in the mentioned plastic
scintillators is only roughly 30\,\%{}, additional means for electron
identification were desirable.    
This was achieved by waveform analysis described in Sections~\ref{sec:meth} and~\ref{sec:elec}.

\end{document}